\documentclass[]{aastex631}

\usepackage{natbib}
\usepackage{color}
\usepackage{xspace}
\usepackage{amsmath}

\newcommand{\radpol}{\texttt{radpol}\xspace}
\newcommand{\ipole}{\texttt{ipole}\xspace}
\newcommand{\grmonty}{\texttt{grmonty}\xspace}

\shorttitle{Compton scattering by accelerated electrons}
\shortauthors{Mo{\'s}cibrodzka}
\graphicspath{{./}{figures/}}

\begin{document}

\title{Polarization-sensitive Compton scattering by accelerated electrons}

\email{E-mail: m.moscibrodzka@astro.ru.nl}

\author[0000-0002-4661-6332]{Monika Mo{\'s}cibrodzka}
\affiliation{Department of Astrophysics/IMAPP, Radboud University,P.O. Box
  9010, 6500 GL Nijmegen, The Netherlands}



\begin{abstract}
We describe upgrades to a numerical code which computes synchrotron and
inverse-Compton emission from relativistic plasma including full polarization.
The introduced upgrades concern scattering kernel which is now capable of
scattering the polarized and unpolarized photons on non-thermal population of 
electrons. We describe the scheme to approach this problem and we test the
numerical code against known analytic solution. Finally, using the upgraded
code, we predict polarization of light that is scattered off sub-relativistic
thermal or relativistic thermal and non-thermal free electrons.
The upgraded code enables more realistic simulations of emissions
from plasma jets associated with accreting compact objects. 
\end{abstract}




\section{Introduction} \label{sec:intro}

Accreting black holes in Active Galactic Nuclei, X-ray
binaries or $\gamma$-ray bursts often produce
relativistic jets. Depending on the system size,
jets are usually observed in radio and infra-red
wavelengths. Interestingly, the radio emission is often
correlated with the X-rays (\citealt{merloni:2003}, \citealt{falcke:2004}). The latter suggests that some of the X-ray
emission observed in accreting black holes may be produced by jets as well.
In such picture, the radio and the X-ray photons are produced by electrons which
experience acceleration.
New insights into black hole accretion and jet emission
may be soon provided by simultaneous spectral-timing-polarimetry at keV energies by missions such as
NASA's X-ray polarimetry mission Imaging X-ray Polarimetry Explorer (IXPE) (\citealt{ixpe:2021}) and
Chinese/European Enhanced X-ray Timing and Polarization mission (eXTP)
(\citealt{zhang:2016}) (and a few other similar experiments). The first
results from IXPE have been recently reported \citep{krawczynski:2022}.
We are therefore motivated to find out what information about electron
acceleration in accretion flows or jets can be carried by polarization of
light, with a particular focus on the inverse-Compton scattered light.

Polarization of X-ray emission (or more generally, higher energy emission) produced by plasma in strong gravity
depends on whether the high energy emission is of synchrotron origin (direct
emission) or arises in the inverse-Compton process (scattered emission).  In
the latter case the polarization of scattered light may be due to transfer of
polarization of synchrotron emission in the inverse-Compton process or may be
due to scattering process itself \citep{chandra:1960}. Hence the polarization of scattered emission depends on many factors: on magnetic field configuration in the
plasma (which impacts polarization of synchrotron radiation),
energy distribution of synchrotron emitting plasma electrons, 
Faraday effects, opacity of the plasma for scatterings or whether the
scattering in the electron frame occurs in Thomson (TH) or Klein-Nishina (KN) regime. In
addition, photon emission and propagation depends on spacetime curvature and on
overall geometry and dynamics of the accretion flow. The complexity of the theoretical
predictions for polarimetric properties of high energy radiation is large (for
complete overview see \citealt{krawczynski:2012}).

To enable theoretical studies of polarimetric properties of emission from complex systems,
we developed \radpol\footnote{{\tt radpol} code is an extention of
  \grmonty which originally assumed unpolarized
emission and emission and scattering off thermal population of electrons
\citep{dolence:2009}. Notice that most of the polarization-insensitive algorithms in \radpol are inherited from \grmonty.} - a covariant Monte
Carlo scheme for calculating multiwavelenght polarized spectral energy distributions (SEDs) of
three-dimentional General Relativistic Magnetohydrodynamic (3-D GRMHD)
simulations of black hole accretion
\citep{moscibrodzka:2020}. The code samples a large
number of polarized synchrotron photons, propagates them in curved spacetime, simulates their
inverse-Compton scatterings and collects information about outgoing spectrum in
a spherically shaped detector at large distance from the center of the model grid.
In our modeling we include synchrotron emission, synchrotron self-absorption in all Stokes parameters and Faraday
effects~\footnote{To integrate radiative transfer equations \radpol is using
  the numerical scheme of another code, \ipole, ray-tracing scheme for
  making polarimetric images of black holes, developed by
  \citet{moscibrodzka:2018}.}, and inverse-Compton process and takes into account all effects
that are important in relativistic plasma in strong gravitational fields of
e.g., black holes. Our method is unique because it is fully covariant
  which enables spectra calculations assuming arbitrary metric tensor.

Our numerical code, until now,
assumed that electron in plasma have thermal distribution function.
In this work we overcome this major over-simplification. 
Here we present a new scattering kernel for \radpol code to permit emission
and polarization from plasma in which electrons are accelerating.
Our model for scattering is completely covariant and allows us to build
more realistic models of emission from relativistic jets.

The structure of the paper is as follows. In Section~\ref{sec:model} we
write basic equations which describe inverse-Compton scattering of polarized
and unpolarized photons off an electron at rest. We then show how scattering is
computed for an ensemble of electrons with four energy distribution functions.
We show that our numerical
method recovers some well known theoretical expectations.
In Section~\ref{sec:results} we present examples of
scattering in Minkowski spacetime that can be used to understand results
from more complex simulations. Section~\ref{sec:diss} list other
code developments carried out to calculate polarized non-thermal spectra of complex accretion
models. We conclude in Section~\ref{sec:con}.

\section{Inverse-Compton scattering model for accelerated electrons}\label{sec:model}

\subsection{\radpol scattering kernel description and upgrades}

We begin with improving the original \radpol polarization-sensitive inverse-Compton
scattering kernel by converting it from {\it an average intensity conserving} one (originally implemented in \radpol) into
{\it a photon conserving} one \citep{schnittman:2013}. The latter make the
scheme more robust and enables us to include scattering off accelerated electrons
with greater precision. 

We first re-consider the inverse-Compton scattering of polarized photon beam in the rest
frame of an electron. The differential cross-section for the Compton scattering
of polarized photons on free electrons is given by the general KN
formula (\citealt{bere:1982}):
\begin{equation}\label{eq:KN_pol}
\frac{d\sigma^{KN}}{d\Omega} = \frac{1}{4} r_e^2
\left(\frac{\epsilon_e'}{\epsilon_e}\right)^2  [ F_{00} 
+  F_{11} \xi_1 \xi_1' + F_1 (\xi_1  + \xi_1')
+  F_{22} \xi_2 \xi_2' + F_{33} \xi_3 \xi_3' ],
\end{equation}
where $r_e=e^2/(4\pi \epsilon_0 m_e c^2)$ is the electron classical radius, $\epsilon_e$ and $\epsilon_e'$
are incident and scattered energy of photon in units of $m_ec^2$,
$\xi_{1,2,3}$ ($\xi_{1,2,3}'$) are normalized polarizations of incident
(scattered) photon, which are defined as follows: $\xi_1 \equiv {\mathcal Q}/{\mathcal I}$, $\xi_2
\equiv {\mathcal U}/{\mathcal I}$, and
$\xi_3 \equiv {\mathcal V}/{\mathcal I}$.
In Equation~\ref{eq:KN_pol}, Stokes ${\mathcal Q}$ and ${\mathcal U}$
(or their fractions $\xi_{1,2}$) are measured with respect to tetrad defined by $\vec{k}$ and the scattering plane, i.e., plane normal to $\vec{k}
\times \vec{k}'$ where $\vec{k}$ ($\vec{k}'$) is an incident (scattered)
photon four-vector in the rest frame of an electron. The coefficients $F$
are elements of the following scattering matrix (\citealt{fano:1949}, \citealt{fano:1957}):
\begin{eqnarray}\label{eq:fano}
{\bf F}=
\frac{1}{2} r_e^2 \left(\frac{\epsilon_e'}{\epsilon_e}\right)^2
\left(
\begin{array}{cccc}
  F_{00} & F_1 & 0 & 0\\
  F_1 & F_{11} & 0 & 0\\
  0   & 0 & F_{22} & 0\\
  0   & 0 & 0 & F_{33}    
\end{array}
\right)= 
\frac{1}{2} r_e^2 \left(\frac{\epsilon_e'}{\epsilon_e}\right)^2
\left(
\begin{array}{cccc}
\frac{\epsilon_e'}{\epsilon_e} + \frac{\epsilon_e}{\epsilon_e'} -\sin^2\theta' &  \sin^2\theta' & 0 & 0  \\
\sin^2\theta' & 1+\cos^2\theta' & 0 & 0\\
 0&0& 2 \cos\theta' &0\\
 0&0&0& \left( \frac{\epsilon_e'}{\epsilon_e} + \frac{\epsilon_e}{\epsilon_e'} \right) \cos \theta'
\end{array}\right)
\end{eqnarray}
where 
$\theta'$ 
is the polar 
scattering angle.
In the TH regime ($\epsilon_e'=\epsilon_e$), $F$ becomes
phase matrix for Rayleigh scattering of Stokes parameters \citep{chandra:1960}. 

Equation~\ref{eq:KN_pol} summed over all possible polarizations of the scattered
photon ($\xi_{123}'$) gives the scattering cross-section as a function of the incident
light linear polarization:
\begin{equation}\label{eq:KN_pol_int}
  \frac{d\sigma^{KN} (\xi_{123}) }{d\Omega} = \frac{1}{2} r_e^2
\left(\frac{\epsilon_e'}{\epsilon_e}\right)^2 
\left( \frac{\epsilon_e}{\epsilon_e'} + \frac{\epsilon_e'}{\epsilon_e} - (1-\xi_1) \sin^2\theta' \right).
\end{equation}
Since $\xi_1$ is defined with respect to scattering plane one can rewrite 
Equation~\ref{eq:KN_pol_int} into:
\begin{equation}\label{eq:KN_pol_int2}
  \frac{d\sigma^{KN} (\xi_{123}) }{d\Omega} = \frac{1}{2} r_e^2
\left(\frac{\epsilon_e'}{\epsilon_e}\right)^2 
\left( \frac{\epsilon_e}{\epsilon_e'} + \frac{\epsilon_e'}{\epsilon_e} -
\sin^2\theta' - \delta \sin^2\theta' cos(2 \phi') \right),
\end{equation}
where $\xi_1={\mathcal Q}/{\mathcal I}= - \delta cos(2\phi')$~\footnote{The minus sign appears because of the
conventions used in this paper and in our numerical code: for fully polarized light,
$\delta=1$, $EVPA=0\deg$ means ${\mathcal Q}$=+1
and $\phi'=90\deg$ measured from x axis, $EVPA=90\deg$ corresponds with
${\mathcal Q}$=-1
and $\phi'=0$ or $180 \deg$.} and where $\phi'$ is the azimuthal scattering angle.
The fractional linear polarization of incident light $\delta = \sqrt{{\mathcal Q}^2+{\mathcal
    U}^2}/{\mathcal I}$ is
invariant to rotations and the azimuthal scattering angle $\phi'$ is measured
with respect to x axis which is chosen arbitrarily.

Sampling of $\theta'$ scattering angle and $\epsilon_e'$ is carried out using azimuthal angle integrated
differential crosssection  and kinematic relation for scattering energy and $\theta'$ angle ($\cos \theta = 1+ 1/\epsilon_e - 1/\epsilon_e'$). This step is polarization independent. 
  
\begin{figure*}[t]\label{fig:StokesI_e}
\def\arraystretch{0.0}
\centering
\includegraphics[width=0.48\linewidth, trim=0cm 15cm 0cm 0cm,clip]{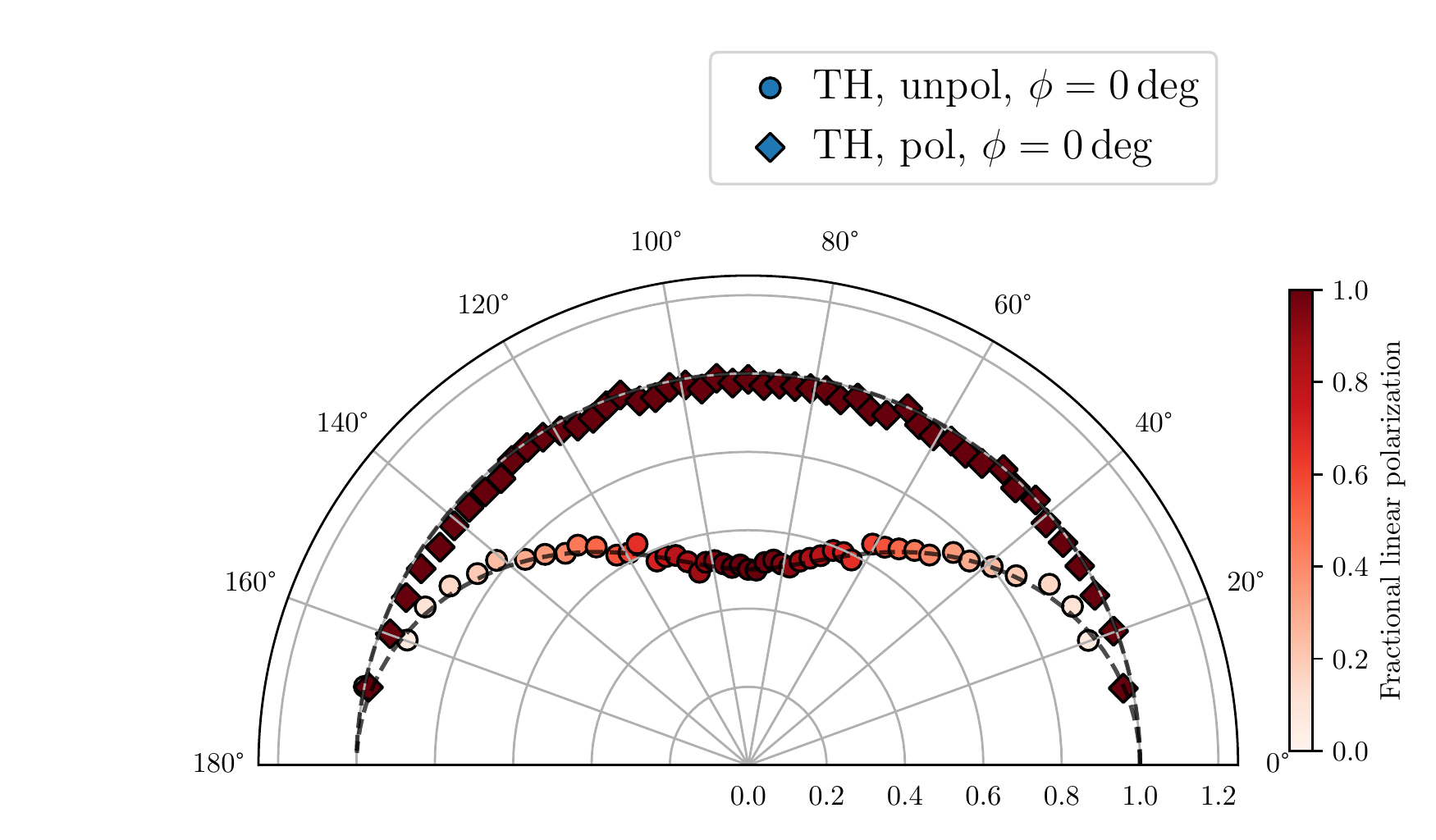}
\includegraphics[width=0.48\linewidth, trim=0cm 15cm 0cm 0cm,clip]{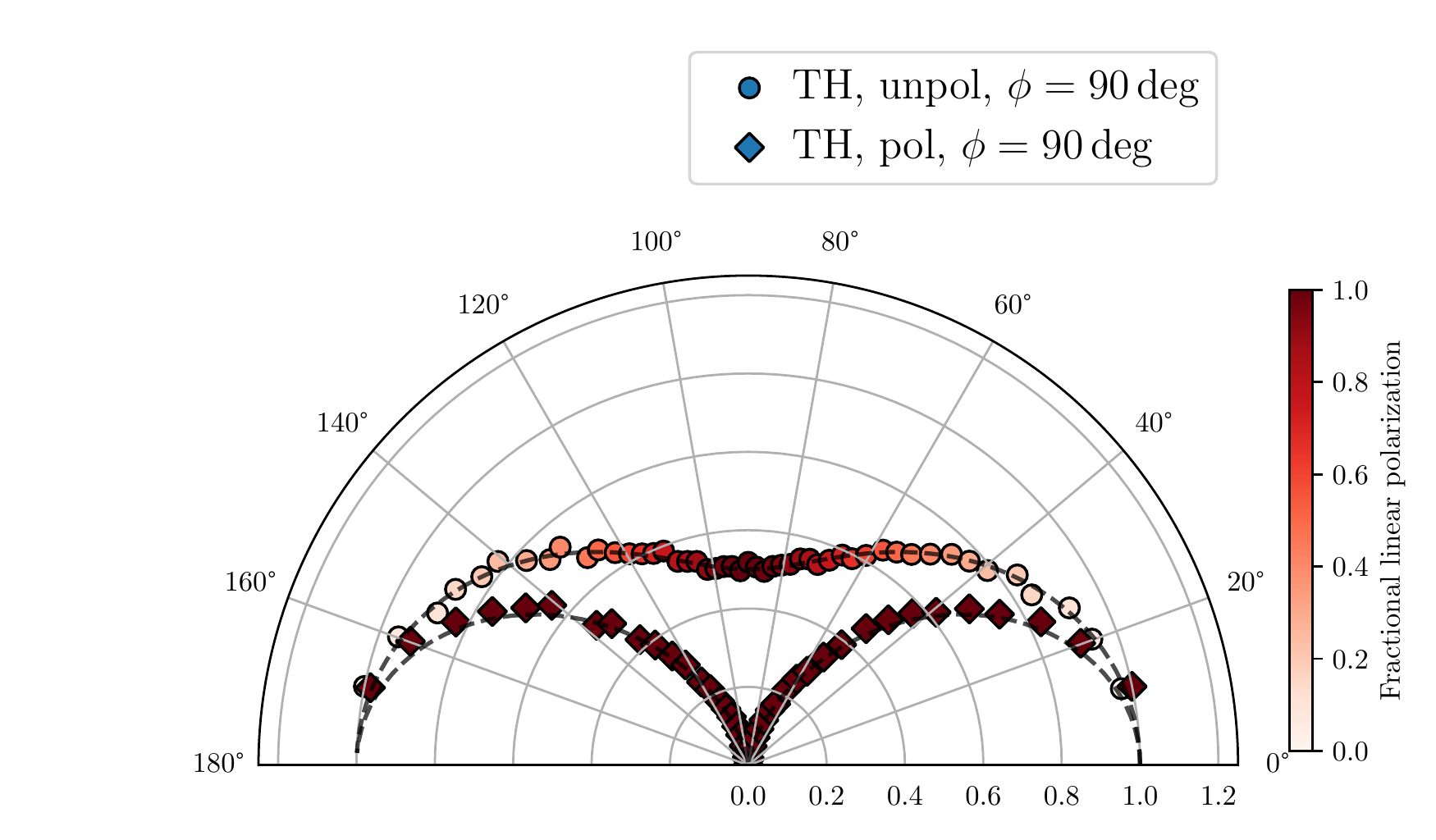}
\caption{Angular histograms showing that our Monte Carlo scheme
  (marked with points) recovers the assumed differential crosssection for
  Compton scattering (marked with dashed lines). Left and right panels display results for azimuthal scattering angles $\phi'=0$ and $\phi'=90^\circ$, respectively. Notice that all angles are measured in the electron rest frame.
The colors encode the scattered light fractional polarizations. When the incident beam is unpolarized (unpol, scattered light marked with circles) then the scattering angle has no azimuthal dependency and light scattered in the direction perpendicular to the incident beam is 100\% polarized (as expected). When the incident beam is fully polarized (pol, scattered light marked with diamonds) the preferred azimuthal scattering angle is that one that is perpendicular to the incident beam polarization direction. Scattered light polarization is then 100\% independently of the scattering angle.}
\end{figure*}  
  
For unpolarized light $\phi' \in (0,2\pi)$ angle can be randomly chosen from a uniform
distribution function, however, if the incident light is polarized, $\phi'$
cannot be random.
The $\phi'$ angle is sampled from the conditional probability distribution function \citep[see][]{zhang:2019}:
\begin{equation}
p(\phi'|\epsilon_e')=\frac{1}{2\pi}-\frac{\delta \sin^2\theta'
  \cos 2\phi'}{2\pi(\frac{\epsilon_e}{\epsilon_e'}+\frac{\epsilon_e'}{\epsilon_e} - \sin^2 \theta')}.
\end{equation}
The $\phi'$ sampling is carried out via inversion of the cumulative
distribution function of the equation above which is:
\begin{equation}
\mathrm{CDF}(\phi')=\frac{\phi'}{2\pi} -\frac{\delta \sin^2\theta
  \sin 2\phi'}{4\pi(\frac{\epsilon_e}{\epsilon_e'}+\frac{\epsilon_e'}{\epsilon_e} - \sin^2 \theta')}
\end{equation}
In the limit of $\delta=0$ or in the limit of $\cos(\theta') = \pm 1$ the formula reduces to
sampling $\phi'$ angle from the uniform distribution. 

Given two scattering angles one can construct $\vec{k}'$ and
define the scattering plane. The fractional Stokes parameters of the scattered photon, $\xi'_{123}$, can be finally computed using:
\begin{eqnarray}\label{eq:fractions}
    \xi_1'=\frac{F_1 + \xi_1 F_{11}}{F_{00}+\xi_1 F_1},\, \,\,
    \xi_2'=\frac{\xi_2 F_{22}}{F_{00}+\xi_1 F_1}, \, \,\,
    \xi_3'=\frac{\xi_3 F_{33}}{F_{00}+\xi_1 F_1}.
\end{eqnarray}
where $\xi_{1,2,3}$ are measured with respect to the scattering plane.
The scattering kernel defined this way is photon-conserving so
Stokes ${\mathcal I}$ does not have to be changed in the scattering
event.
In the originally published version of \radpol, we sampled $\phi'$
angle from uniform distribution function so transformation of polarization
included transformation of all Stokes
parameters, including Stokes ${\mathcal I}$, using Equation~\ref{eq:fano}.
Hence, the original scheme was not photon conserving but only averaged intensity
conserving \citep{schnittman:2013}. 

We have tested the new implementation of the Compton scattering in electron rest-frame.
If we reconsider scattering of photons in the electron rest-frame, the scattering angle $\phi'$ depends on the
polarization degree and angle of the incident light. For fully polarized
light, i.e., $\delta=1$, the scattering of polarized light is favored in the
direction perpendicular to the polarization angle.
In Figure~\ref{fig:StokesI_e} we show that the outcome of
our numerical calculations are consistent with these theoretical expectations
(marked in the figure as dashed line). Scattering an unpolarized light
  off an electron at rest can produce polarized emission for scattering angles $\theta'= 90 \deg$.

\subsection{Electron Acceleration Models}\label{sec:eDF}

Next we consider scattering off a population of electrons.
We assume the following electron energy distribution functions (eDFs) that are
usually considered for astrophysical applications.

\begin{itemize}
\item relativistic thermal eDF (\citealt{Petrosian:1981}, \citealt{Leung:2011}):
\begin{equation}
\frac{1}{n_e}\frac{dn_e}{d\gamma} = \frac{\gamma^2 \beta } {\Theta_e K_2(1/\Theta_e)} \exp^{(-\gamma/\Theta_e)}
\end{equation}
where $\beta \equiv \sqrt{1-1/\gamma^2}$ and
$\Theta_e=k_b T_e/m_e c^2$ is the dimensionless electron temperature,

\item purely power-law eDF (\citealt{RL:1979}):
\begin{equation}
\frac{1}{n_e}\frac{d n_e}{d\gamma} = \frac{ (p-1)}{(\gamma_{min}^{1-p} - \gamma_{max}^{1-p})} \gamma^{-p}
\end{equation}
where $p$, $\eta$, $\gamma_{min}$ and $\gamma_{max}$ are parameters,

\item hybrid eDF where we assume that the electrons are accelerated from
 a thermal eDF. Accelerated electrons energies are described by a power-law distribution:
\begin{equation}
\frac{1}{n_{pl}}\frac{d n_{pl}}{d\gamma}=\frac{(p-1)}{(\gamma_{min}^{1-p} - \gamma_{max}^{1-p})} \gamma^{-p},
\end{equation}
where $\gamma_{\rm min}$, $\gamma_{\rm max}$, and $p$ are parameters of the
acceleration model (we will assume that $\gamma_{\rm max} \gg 1$, in practice
we assume $\gamma_{\rm max}=10^6$). The power-law function is ``stitched'' to
the thermal eDF as follows (the same methodology is presented by
\citealt{ozel:2000} and \citealt{yuan:2003}).
The energy density of the thermal electrons is
\begin{equation}\label{eq:uth}
  u_{th}=n_{th} \Theta_e a(\Theta_e) m_e c^2
\end{equation}
where $a(\Theta_e)\approx(6+15\Theta_e)/(4+5\Theta_e)$ \citep{gammie:1998} while the energy density of the accelerated electrons is
\begin{equation}\label{eq:upl}
  u_{pl}=n_{pl} \frac{p-1}{p-2} \gamma_{min} m_e c^2.
\end{equation}
where the simple form of $u_{pl}$ is due to normalization of the power-law function. 
We assume that $u_{pl}=\eta u_{th}$ where $\eta$ is a fourth free parameter of
the acceleration model indicating the fraction of thermal energy transferred to
the non-thermal tail. Using Equations~\ref{eq:uth} and~\ref{eq:upl} we
calculate the resulting number density of accelerated electrons, $n_{pl}$:
\begin{equation}\label{eq:factorf}
n_{pl}=\frac{p-2}{p-1} \gamma_{min}^{-1} \eta a(\Theta_e) \Theta_e n_{th}.
\end{equation}
In this model the power-law eDF should smoothly connect with
the thermal distribution so we require that:
\begin{equation}
n_{th}(\gamma_{min})=n_{pl}(\gamma_{min}).
\end{equation}
For a set of $p$, $\eta$ and $\Theta_e$, we solve 
\begin{equation}\label{eq:gammin}
\gamma_{min}^4 \beta_{min} \exp(-\gamma_{min}/\Theta_e)= 2 (p-2) \eta a(\Theta_e) \Theta_e^4
\end{equation}
to find the $\gamma_{\rm min}$.

\item $\kappa$ eDF (\citealt{Vas:1968}, \citealt{Xiao:2006}, \citealt{Pierrard:2010}) is a more natural eDF inspired by kinetic studies
  of relativistic plasmas:
\begin{equation}
\frac{1}{n_e} \frac{d n_e}{d\gamma}= \gamma \sqrt{\gamma^2-1} \left(1+\frac{\gamma+1}{\kappa w}\right)^{-(\kappa+1)}
\end{equation}
where $\kappa$ and $w$ are parameters. For $\kappa \rightarrow \infty$,
$\kappa$ distribution function becomes Maxwell-J{\"u}ttner distribution.

\end{itemize}

\subsection{Thermal and non-thermal electron energy sampling}\label{app:nonth}

In upgraded \radpol, the scattering kernel is sampling electron four momentum
$p^\mu$ from thermal and non-thermal distribution functions above assuming
that the spatial parts of electron four-momentum are isotropic in the fluid
co-moving frame. Isotropic eDF model limits the discussion to energy sampling.

To sample electron Lorentz factor $\gamma_e$ in thermal distribution function we use the
sampling procedure introduced by \citet{canfield:1987} (implemented in \grmonty
and \radpol codes).

In case of pure power-law distribution function the electron Lorentz factor is
sampled using inversion of cumulative distribution function where the
inversion has analytic form:
\begin{equation}
  \gamma_e= \left(\gamma_{min}^{1-p}(1-r) + \gamma_{max}^{1-p}r \right)^{1/(1-p)}
\end{equation}
where $r\in(0,1)$ is a random number and $\gamma_{min},\gamma_{max},p$ are the eDF parameters.

To sample Lorentz factor from hybrid and $\kappa$ distribution functions we
re-write these two eDFs as a product of two
probability functions $p_1$
and $p_2$, where $p_1$ is used for tentative sampling and $p_2$ is used for rejection sampling
(the procedure closely follows \citealt{canfield:1987} but differs in details
of tentative sampling).
For both hybrid and $\kappa$ DF:
\begin{equation}
p_1=\frac{1}{n_e} \frac{dn_e(\gamma)}{d\gamma_e} \frac{1}{\beta_e}
\end{equation}
and
\begin{equation}
p_2=\beta_e
\end{equation}
where $\beta_e=\sqrt{1-1/\gamma_e^2}$.

In our model the tentative sampling of $\gamma_e$ from $p_1$ is carried out by inversion of cumulative distribution
function. We found analytic forms of cumulative distribution function of $p_1$ (hereafter modified cumulative distribution function, MCDF) for
hybrid and $\kappa$ eDFs. For hybrid distribution function it is:
\begin{eqnarray}
  \mathrm{MCDF}_{\rm hybrid}(\gamma_e)= 1-\frac{\exp{(-\frac{\gamma_e}{\Theta_e})}}{\exp(-\frac{1}{\Theta_e})}
  \frac{(2\Theta_e^2+2\Theta_e\gamma_e+\gamma_e^2)}{(2\Theta_e^2+2\Theta_e+1)}  (1-f) +
  \\
  \begin{cases}
    0 & \mathrm{for} \,\, \gamma_e<\gamma_{min} \\
    f \frac{(p-1)}{(\gamma_{min}^{1-p}-\gamma_{max}^{1-p})}
    \left(g_{pl}(\gamma_e,p)-g_{pl}(\gamma_{min},p)\right) & \mathrm{for} \,\,\gamma_e>\gamma_{min}
\end{cases}
\end{eqnarray}
where the third term is added only for $\gamma_e>\gamma_{min}$, where
$f=n_{pl}/n_{th}$ (given by Equation~\ref{eq:factorf}) and where:
\begin{eqnarray}
  g_{pl}(\gamma_e)=
\begin{cases}
  \sqrt{\gamma_e^2-1}\left(\frac{1}{\gamma_e}\right) & \mathrm{for} \,p=3 \\
  \sqrt{\gamma_e^2-1}\left(\frac{1}{2\gamma_e^2}\right)-\frac{1}{2}\arcsin(\frac{1}{\gamma_e}) & \mathrm{for} \,p=4 \\
  \sqrt{\gamma_e^2-1}\left(\frac{2}{3\gamma_e}+\frac{1}{3\gamma_e^3}\right) & \mathrm{for} \,p=5 \\
  \sqrt{\gamma_e^2-1}\left(\frac{3}{8\gamma_e^2}+\frac{1}{4\gamma_e^4}\right)-\frac{3}{8}\arcsin(\frac{1}{\gamma_e}) & \mathrm{for} \,p=6.
\end{cases}
\end{eqnarray}
For $\kappa$ eDF the $p_1$ cumulative distribution
function for sampling $\gamma_e$ is:
\begin{eqnarray}
  \mathrm{MCDF}_{\kappa}(\gamma_e)= f_{\kappa,n}
  \left( f_{\kappa,1} e^{(\kappa log(\gamma_e+\kappa w+1))}+
  f_{\kappa,2} e^{(\kappa log(\kappa w+2))}\right)/(\kappa^2-3\kappa+2)\\
    e^{(-\kappa \log(\gamma_e+\kappa w+1) -\kappa \log(\kappa w+2))}
\end{eqnarray}
where 
\begin{eqnarray}
    f_{\kappa,1}=w^{\kappa+1} \left(2\kappa^{\kappa+2}w^2+
    (2\kappa^{\kappa+2}+4 \kappa^{\kappa+1})w
    +(\kappa^{\kappa+2}+\kappa^{\kappa+1}+2 \kappa^\kappa)\right),\\
    f_{\kappa,2}=\kappa^\kappa (\kappa-\kappa^2) w^{\kappa+1} \gamma_e^2+ 
    w^{\kappa} (-2\kappa^{\kappa+2}w^2 - 2 \kappa^{\kappa+1}w) \gamma_e+\\
    w^{\kappa} (-2\kappa^{\kappa+2}w^3 - 4 \kappa^{\kappa+1}w^2-2\kappa^\kappa
    w).
\end{eqnarray}
and the distribution normalizing factor $f_{\kappa,n}$ is given in
\citet{pandya:2016} (see their Equation 19).
For fast and accurate numerical MCDF inversion we use Regula-Falsi 
root finder \citep{ford:1995}. Since $\beta$ is close to one for relativistic electrons the
rejection sampling is efficient.

\subsection{Test of the numerical scheme against analytic model}

\begin{figure*}
\def\arraystretch{0.0}
\centering
\includegraphics[width=0.4\textwidth,angle=0]{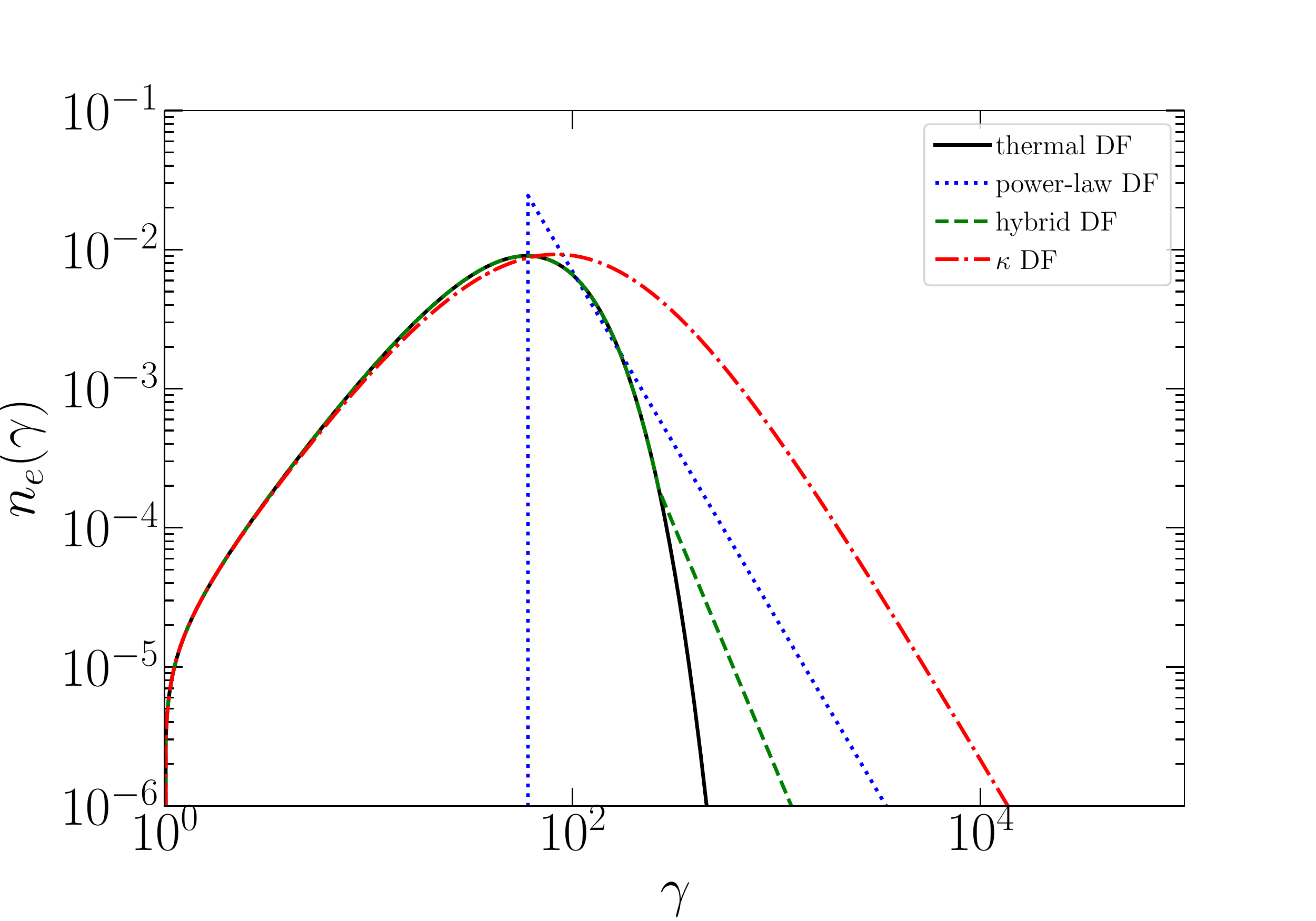}
\includegraphics[width=0.4\linewidth]{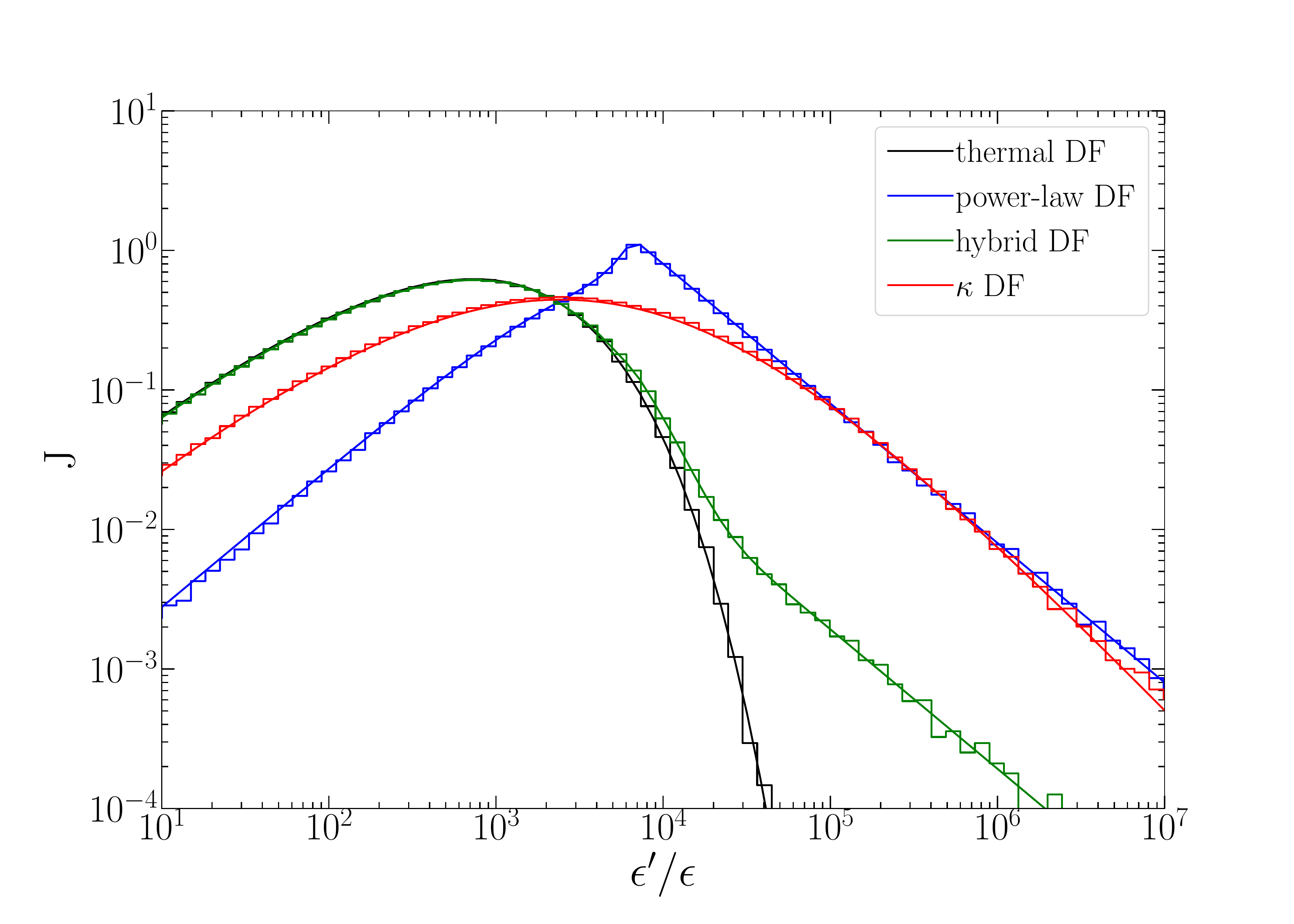}\\
\includegraphics[width=0.4\linewidth]{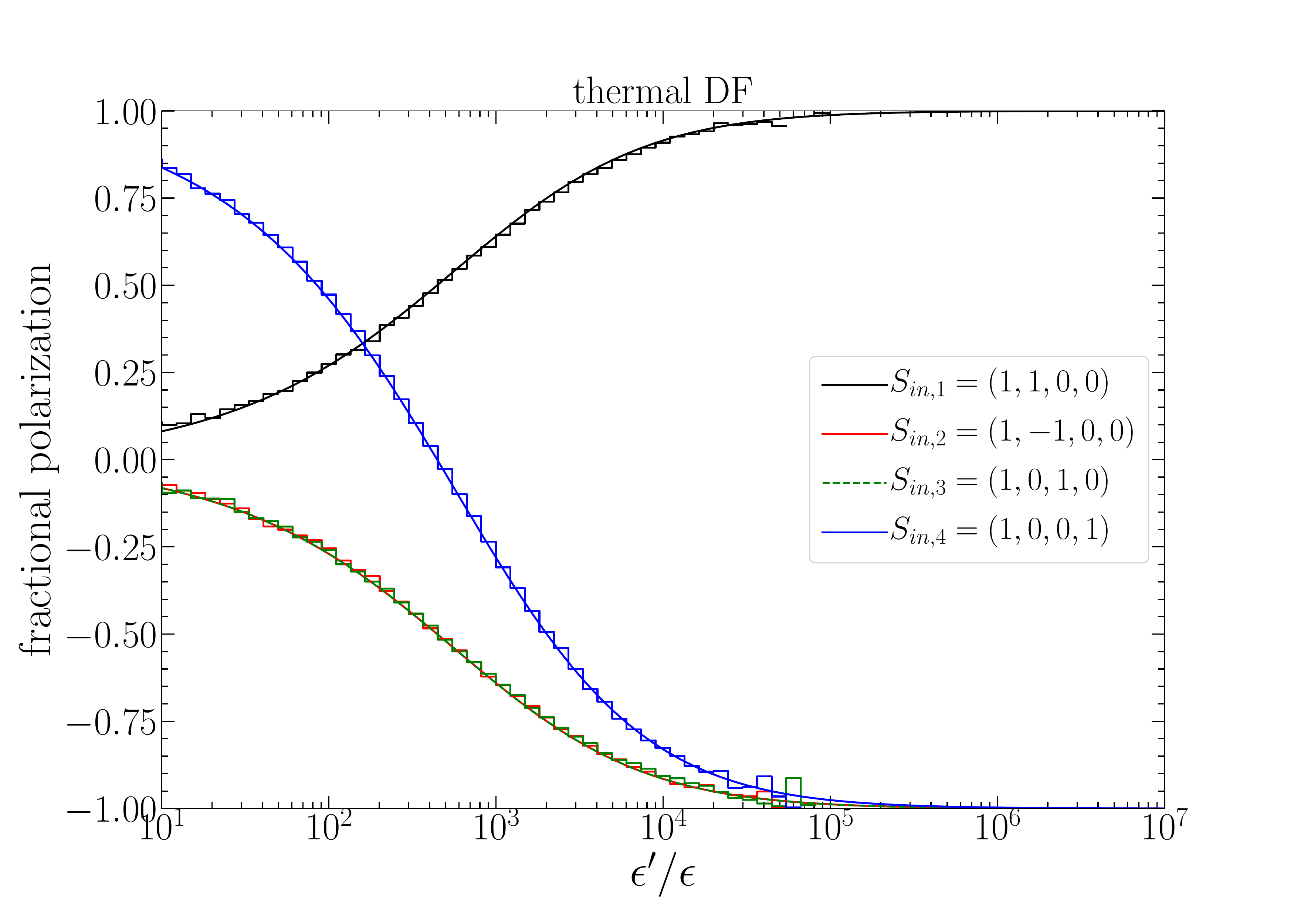}
\includegraphics[width=0.4\linewidth]{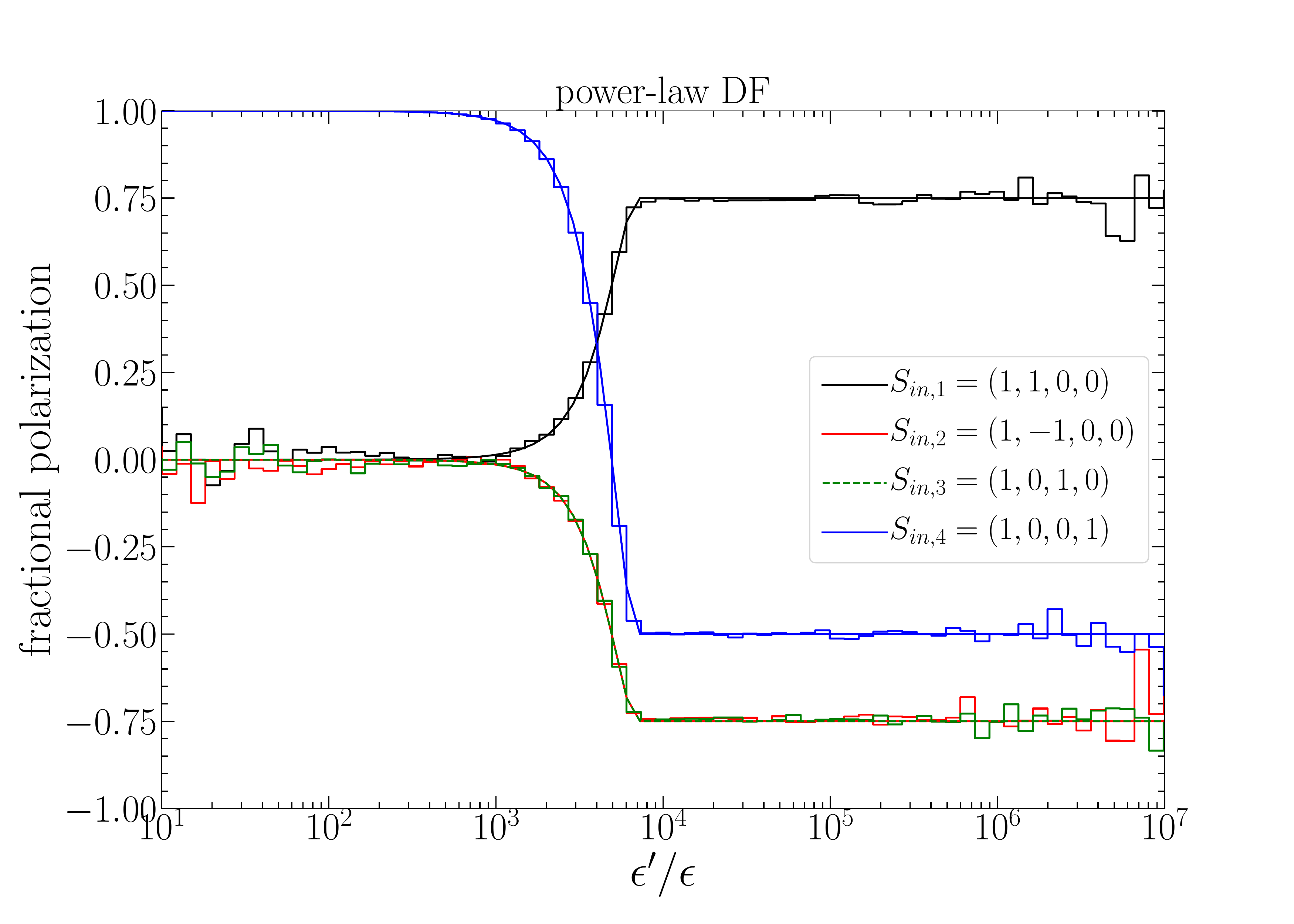}\\
\includegraphics[width=0.4\linewidth]{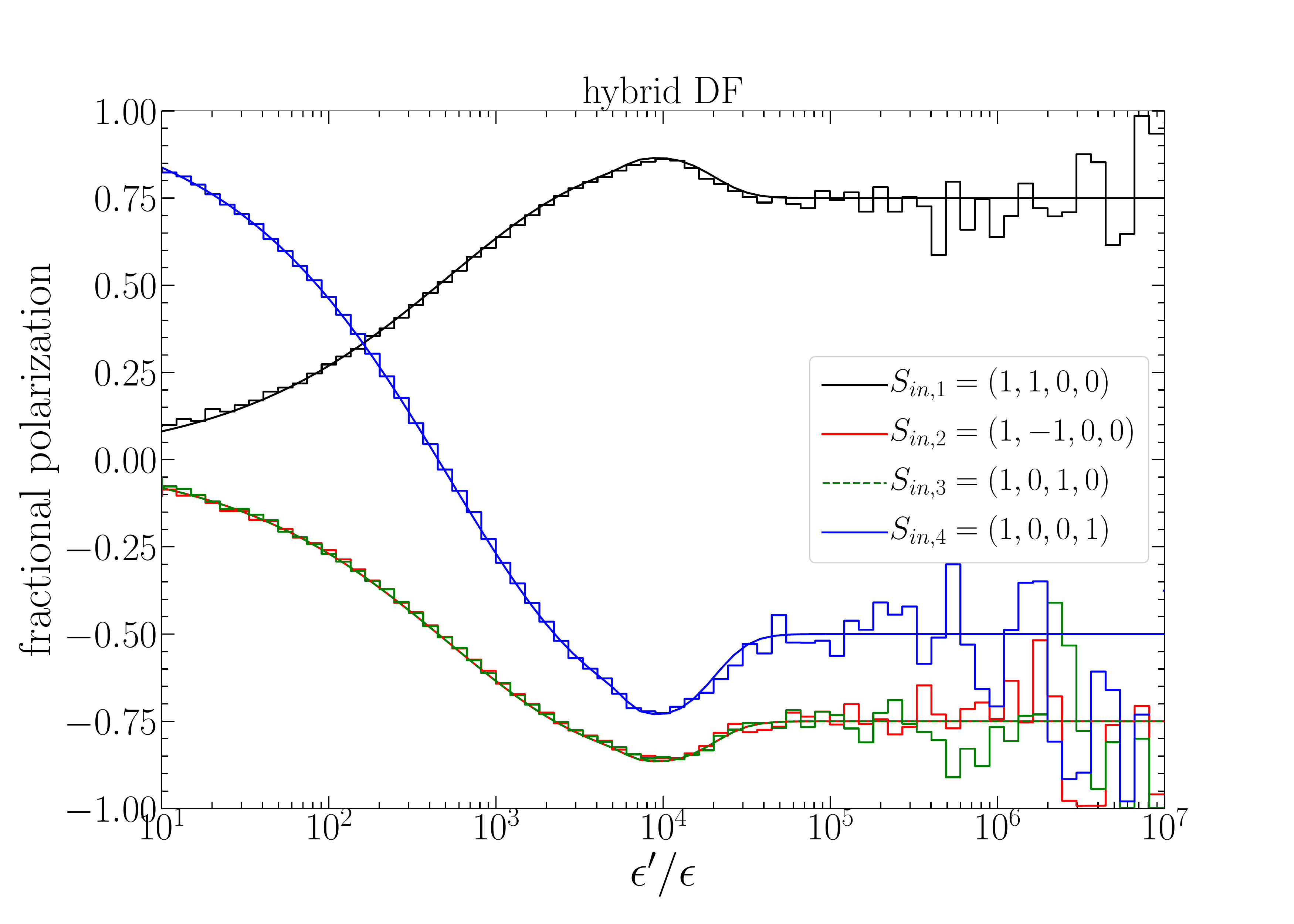}
\includegraphics[width=0.4\linewidth]{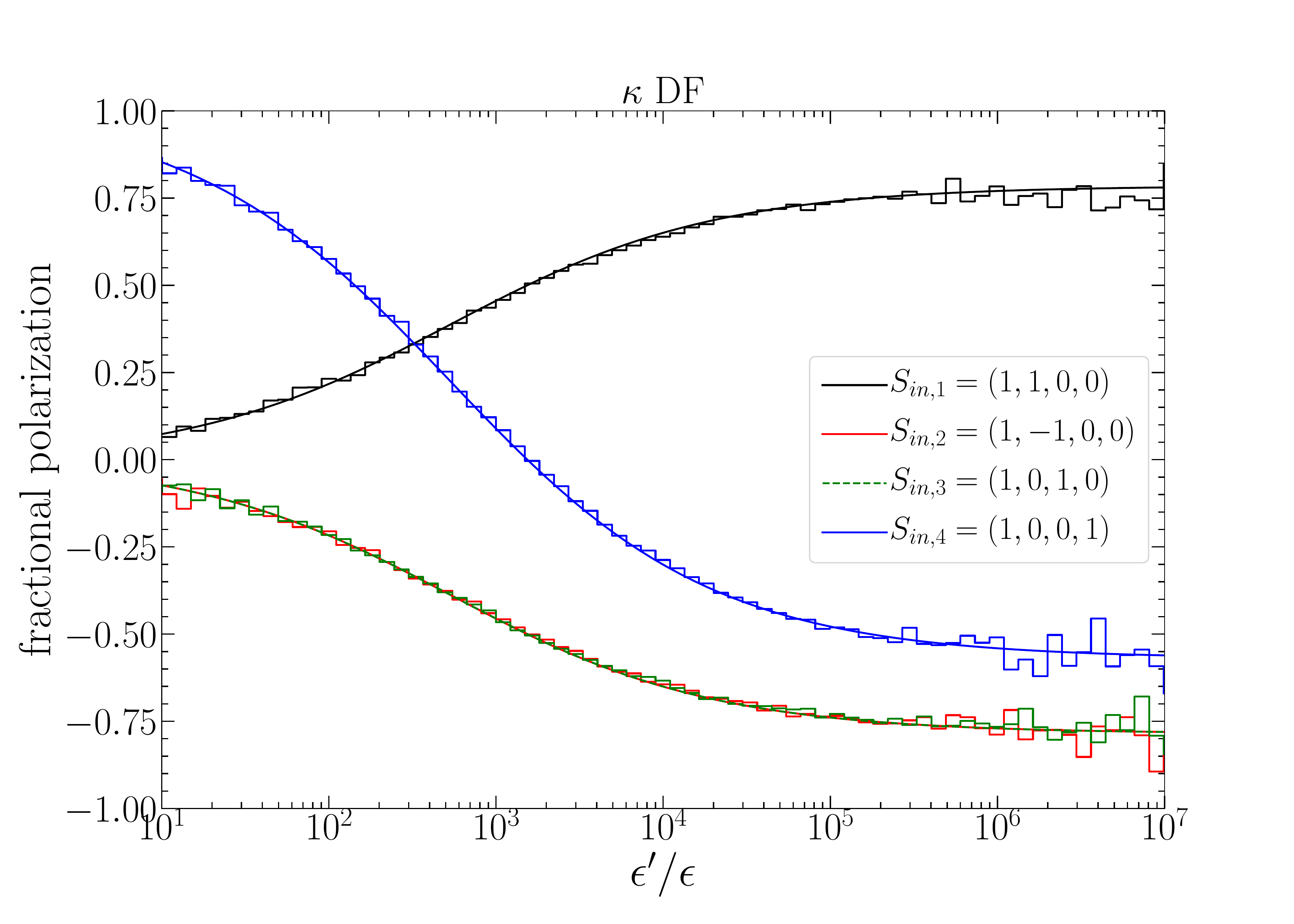}
\caption{Comparison of our Monte Carlo scattering kernel results to analytic expectation for scattering angle of $\theta'=85^circ$ (measured in the laboratory with respect to the direction of the incident beam). Panels show: the electron distribution function models (top left panel); the scattered light intensity (top right panels) and light polarizations (middle and bottom panels). Panels with fractional polarizations show results as a function of the incident photon polarization state, $S_{\rm in}$. The thermal eDF model assumed
  electron dimensionless temperature $\Theta_e=k_B T_e /m_e c^2 =
  10$. In purely power-law model: $p=3$, $\gamma_{\rm min}=60$, $\gamma_{\rm max}=10^6$. In the hybrid model we assume the following parameters of the non-thermal tail: $p=3$, $\eta=0.1$, $\gamma_{\rm max}=10^6$ and $\gamma_{\rm min}$ is found by solving Equation~\ref{eq:gammin}.  
The $\kappa$ eDF parameters are: $\kappa=4.5$ and w=10.}\label{fig:StokesI_e2}
\end{figure*}

To test numerical code we consider single scattering of a beam of monochromatic polarized photons off
an enable of electrons with four eDFs introduced in the previous sub-sections.  \citet{bonometto:1970} provided semi-analytic solution to
this problem as long as electron-frame scattering is in TH limit ($\epsilon'=\epsilon$). The analytic model has been already briefly described in Appendix A of our
previous work \citep{moscibrodzka:2020} and recently also reproduced in more
details by \citet{yang:2021}.

Our numerical model can be confronted with the theoretical expectation for light intensity and polarization with predictions of \citet{bonometto:1970}.  
In Figure~\ref{fig:StokesI_e2} we show
agreement between the theoretical prediction with our numerical kernel calculations 
using our new updated scattering kernel using thermal, power-law, hybrid
and $\kappa$ electron distribution functions for single scattering angle.
The Monte Carlo simulations with \radpol scattering kernel converge to the predicted values. Our results are also consistent with results presented in 
\citealt{yang:2021} (see their Figure 25) who carried out the same tests using independent numerical scheme. 
In all cases the fractional linear polarization is increasing with
frequency. In particular, for eDF with a power-law component (power-law, hybrid, and $\kappa$
eDFs) the fractional linear polarization converges to a constant value at high energies ($\epsilon' \gg \epsilon$) in analogy 
to the fractional linear polarization of the optically thin synchrotron emission (which can be also thought of as a scattering process) from electrons distributed into a power-law eDF.

\section{Scattering off low- and high-energy thermal and
  non-thermal electrons}\label{sec:results}

\begin{figure*}
\def\arraystretch{0.0}
\centering
\setlength{\tabcolsep}{0pt}
\begin{tabular}{ccc}
{\bf \,\, Maxwell-J{\"u}ttner eDF $\Theta_e$=0.1} & {\bf \,\, Maxwell-J{\"u}ttner eDF $\Theta_e$=100}  &
  {\bf Nonthermal $\kappa$ eDF}\\
\includegraphics[width=0.33\linewidth]{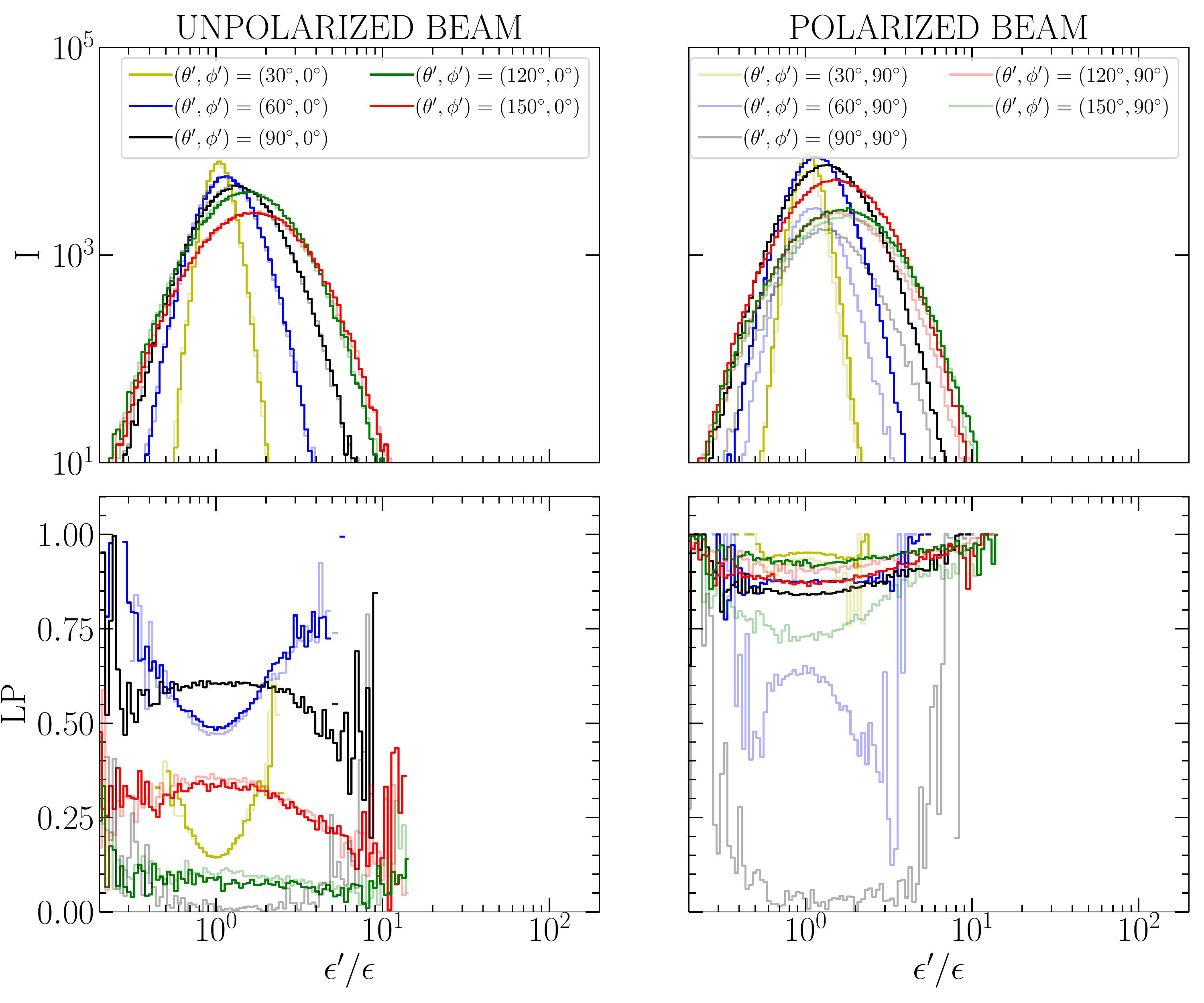} &
\includegraphics[width=0.33\linewidth]{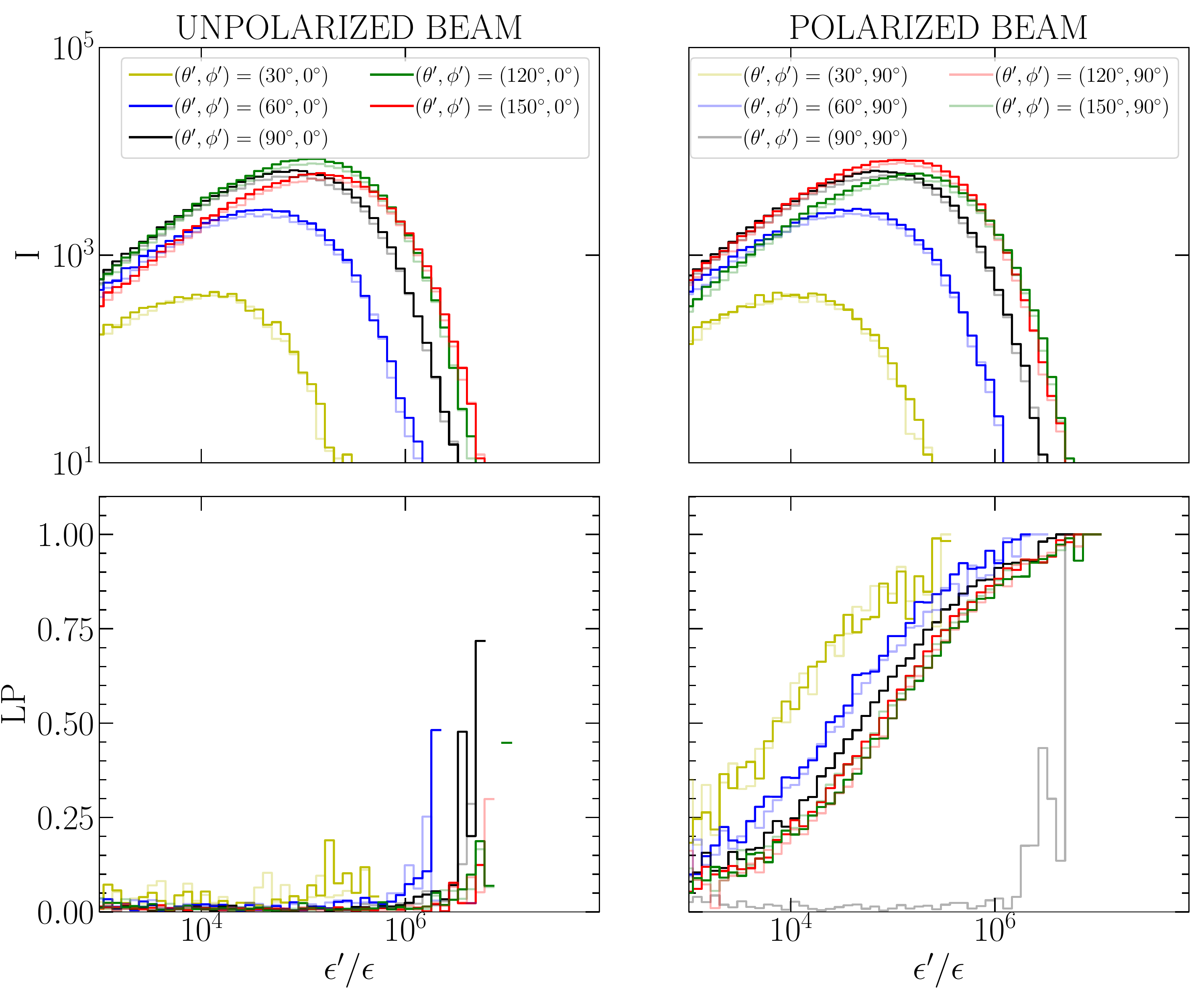} &
\includegraphics[width=0.33\linewidth]{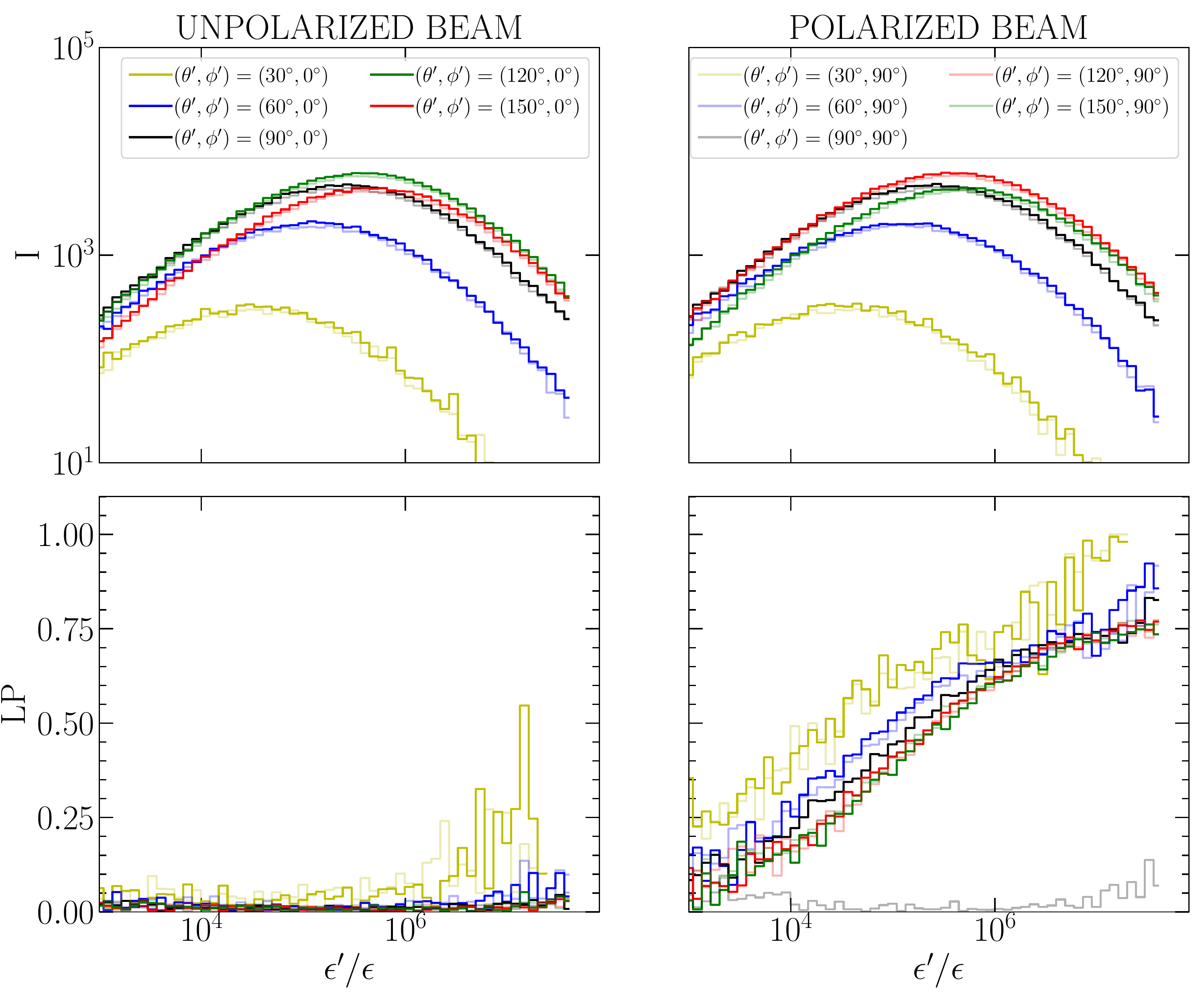}
\end{tabular}
\caption{SEDs of light scattered on electrons with thermal (sub-relativistic and relativistic plasma)
  and non-thermal energy distribution functions. Upper panels display
  the scattered light intensity (I) and lower panels show the scattered light
  fractional linear polarization (LP). Line colors and transparency encode the directions of scattering with respect to the incident photon beam in the laboratory frame. The model assumes single TH scattering of a monochromatic beam.}\label{fig:final_TH}
\end{figure*}

\begin{figure*}
\def\arraystretch{0.0}
\centering
\setlength{\tabcolsep}{0pt}
\begin{tabular}{ccc}
{\bf \,\, Maxwell-J{\"u}ttner eDF $\Theta_e$=0.1} & {\bf \,\,  Maxwell-J{\"u}ttner eDF $\Theta_e$=100}  &
  {\bf Nonthermal $\kappa$ eDF}\\
\includegraphics[width=0.33\linewidth]{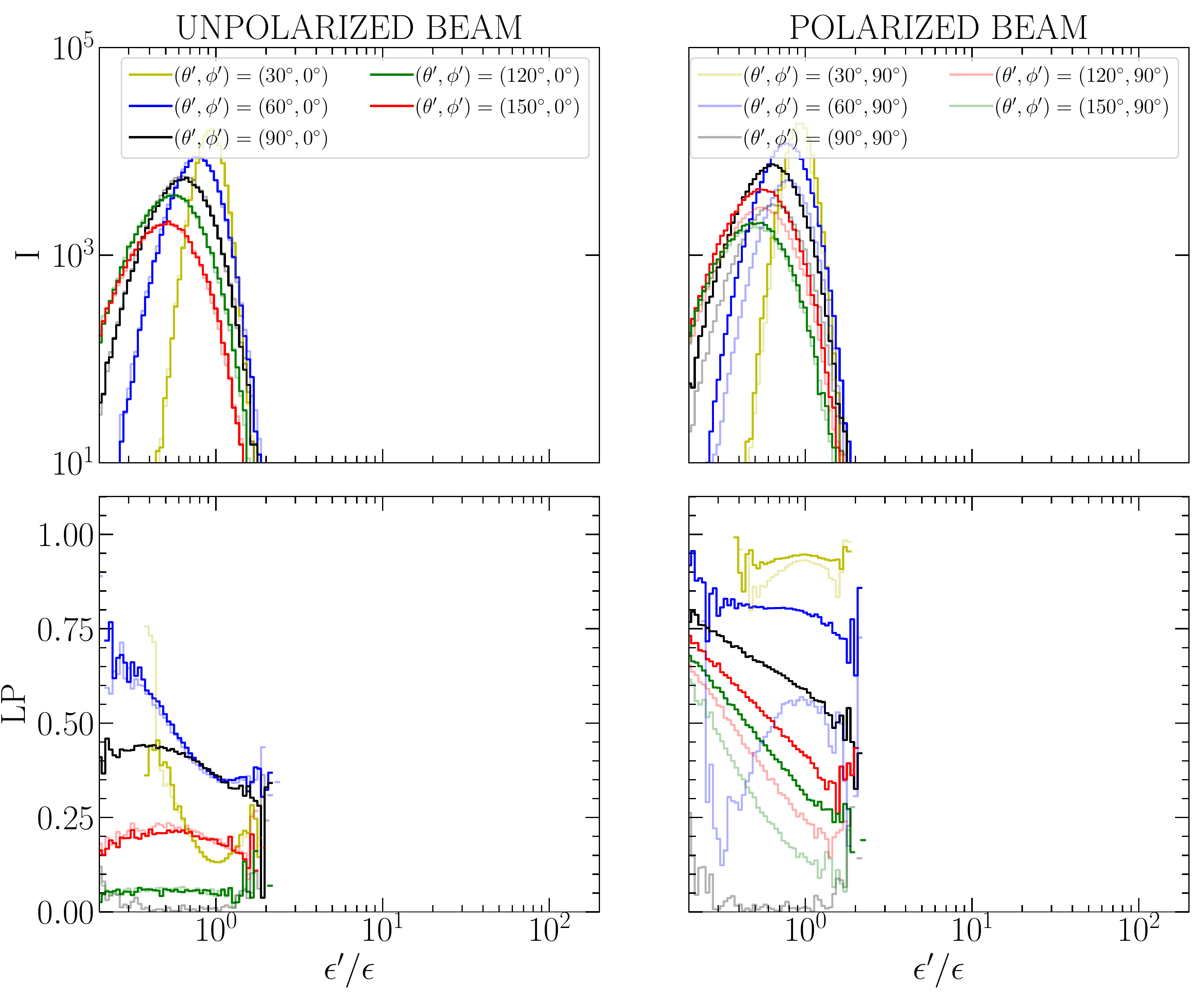} &
\includegraphics[width=0.33\linewidth]{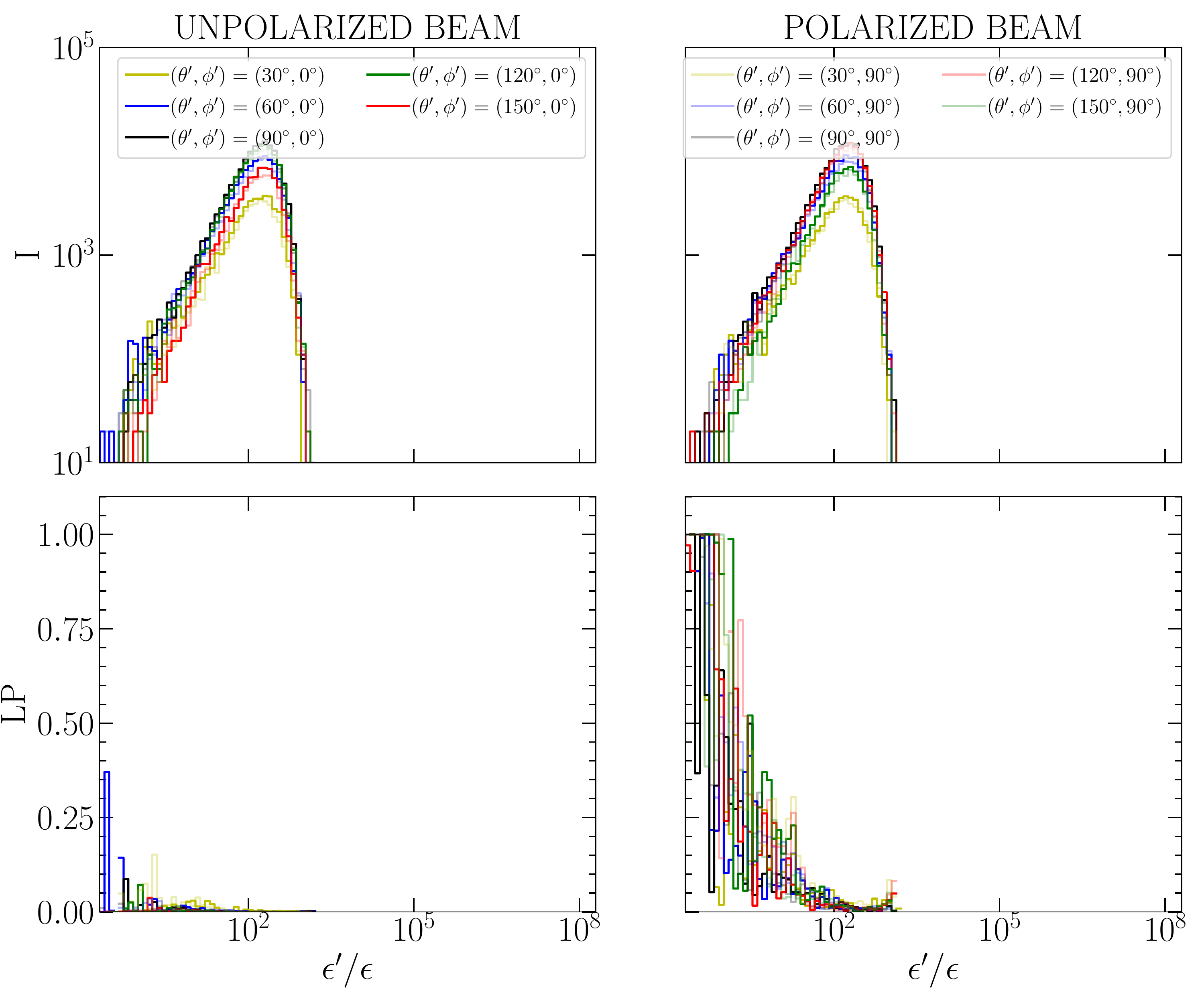} &
\includegraphics[width=0.33\linewidth]{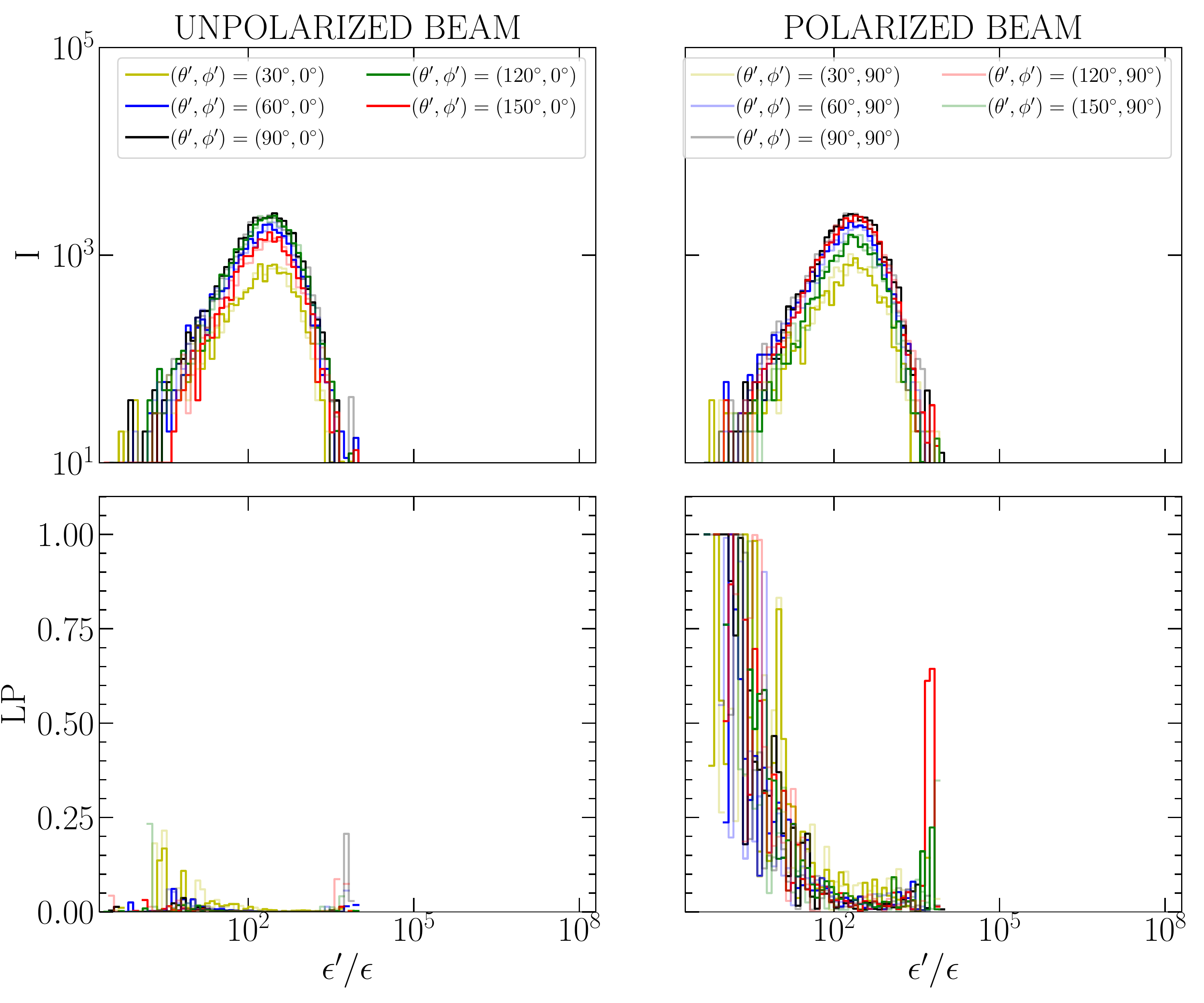}
\end{tabular}
\caption{Same as in Figure~\ref{fig:final_TH} but for scattering in
  KN regime.}\label{fig:final_KN}
\end{figure*}

Next we simulate a single inverse-Compton scattering of monochromatic beam as
a function of the incident light polarization, eDF, and scattering regime (TH and KN). It is expected that scattering of unpolarized photon beam of hot relativistic plasma should produce no polarization (e.g., \citealt{poutanen:1993}), here we can test our code against this expectation. Otherwise, the results presented in this section can be used as a guiding line for analysis of more complex models (e.g., radiation produced in accretion disks and jets in GRMHD simulations), keeping in mind that in realistic accretion flows and jets scatterings may be multiple. 
Notice that here we neglect circular polarization of the incident beam because the circular polarization cannot be generated in the scattering process.

In Figure~\ref{fig:final_TH} we show intensity (upper panels) and fractional polarization (lower panels) spectra of scattered
light when the scattering occurs in the TH regime (i.e. the energy of the
incident beam is low compared to the electron rest mass energy,
$\epsilon = 2.5\times10^{-11}$). Panels left to right display results for scattering on
sub-relativistic (characterized by the dimensionless temperature
$\Theta_e=0.1$) and relativistic electrons distributed into thermal (with
$\Theta_e=100$) and $\kappa$ (with $w=100$ and $\kappa=4.5$) eDF.
Initially unpolarized light ($S_{in}=(1,0,0,0)$) scattering off an ensemble of subrelativistic electrons
becomes polarized and the degree of polarization depends on the geometry (angle) of scattering
and on the scattered photons frequency.
Initially polarized light ($S_{in}=(1,1,0,0)$) scattering off cold electrons will
stay polarized only for certain scattering angles. 
Scattering unpolarized beam off hot (relativistic) electrons does not produce polarization, as expected. (The
residual polarization seen in the high energies in this case is a Monte Carlo noise.)
The latter is valid for thermal and non-thermal electron distribution function. For initially polarized beam
scattering off relativistic electrons, the scattered radiation is partially polarized
with fractional polarization increasing with frequency from zero to 100 \%. Only for a very specific
scattering angle ($(\theta',\phi')=(90^\circ,90^\circ)$) the polarization
cancels out to zero.

In Figure~\ref{fig:final_KN} we display results of the same numerical tests as
shown in Figure~\ref{fig:final_TH} but with scatterings in KN
regime (i.e. the energy of the
incident beam is comparable to the electron rest mass energy,
$\epsilon = 1$). Scattering off cold electrons produces variety of
  polarizations which depend on the scattering direction, similar to
  results for scattering in the TH regime. For KN scattering
  off relativistic electrons, the initially unpolarized light will not gain
  any polarization independently of the eDF, consistent with results in the TH regime. However, for incident polarized light the polarization of
  the scattered light is sharply decreasing with frequency regardless of the
  scattering angle which is the opposite trend compared to the TH scattering. Noteworthy, as evident in both Figures~\ref{fig:final_TH} and~\ref{fig:final_KN}, the total intensity of the scattered light slightly
  depends on the incident light polarization.

\section{Polarimetric properties of scattered light in complex models of
  accretion}\label{sec:diss}

Our upgraded scattering kernel in \radpol code is now well tested
and produces results consistent with theoretical expectations for variety of
electron distribution functions. 
Simulating polarized emission and scattering off non-thermal electrons in
complex models of accretion (for example in GRMHD simulations of accreting
black holes) requires modifications of the photon sampling routines as well as
scattering cross-sections. Manufacturing photons in
\radpol is carried out just like in its
unpolarized version \grmonty (see method paper by \citealt{dolence:2009}) with a difference that
now all angle averaged synchrotron
emissivities incorporate thermal and non-thermal eDF. 
Once a photon wavevector, $k^\mu$, is build in the fluid frame,
the photon polarization is assigned to it using corresponding thermal/non-thermal synchrotron emissivities.
Finally, to determine the place of scattering along a
ray path in \radpol simulation, an optical depth for scattering is calculated in each step on
geodesic path. The so called ``hot crosssection'' is calculated to estimate cross-section for a photon
interaction with an ensemble of free electrons. This requires integrating
KN (or TH) cross-section over assumed electron distribution function that
can be now also non-thermal. In \radpol such integrations are done numerically and tabulated. 
Full exploration of polarization of high energy emission produced in complex
models of accretion flows with electron acceleration is beyond the scope of this work and will be
presented in the forthcoming publication.

\section{Conclusion}\label{sec:con}

In \citet{moscibrodzka:2020} we have introduced a Monte Carlo code \radpol, which is
capable of tracing light polarization of synchrotron emission and
polarization-sensitive inverse-Compton scattering processes in full general relativity. 
In the current work we describe a major extension of the code to compute
emission and scattering when electrons are non-thermal. 
The numerical scheme tests converge to the theoretical expectations.
Updated code enables more realistic fully relativistic and covariant models of
emission for jets produced by accreting objects of any kind.

\section*{Acknowledgements}
The author thanks Hector Olivares for comments on Regula-Falsi root finder.
The author acknowledges support by the NWO grant no. OCENW.KLEIN.113.

\bibliography{local}{}
\bibliographystyle{aasjournal}



\end{document}